\documentclass[showpacs,twocolumn,prl]{revtex4}
\usepackage{amsmath,graphicx}
\usepackage{textcomp}
\usepackage{amssymb}
\usepackage{comment}
\usepackage[bookmarks=false,linkcolor=blue,urlcolor=blue,colorlinks,citecolor=blue]{hyperref}
\usepackage[normalem]{ ulem } 
\usepackage{soul}             
\usepackage[svgnames]{xcolor}

\definecolor{Yangcolor}{RGB}{1, 148, 109}

\begin{document}

\title{Long-lived Andreev states as evidence for protected hinge modes in a bismuth nanoring Josephson junction}
\author{A.~Bernard$^{1}$, Y.~Peng$^{2,3}$,  A.~Kasumov$^{1,4}$, R. Deblock$^{1}$, M. Ferrier$^{1}$, F. Fortuna$^{5}$, V.~T.~Volkov$^{4}$, Yu.~A.~Kasumov$^{4}$, Y. Oreg$^{6}$, F.~von~Oppen$^{7}$, H.~Bouchiat$^{1}$ and S.~Gu\'eron $^{1}$ }
		\affiliation{$^{1}$ Universit\'e Paris-Saclay, CNRS, Laboratoire de Physique des Solides, 91405, Orsay, France.}
  \affiliation{$^{2}$Department of Physics and Astronomy, California State University, Northridge, California 91330, USA}
  \affiliation{$^{3}$Institute of Quantum Information and Matter and Department of Physics, California Institute of Technology, Pasadena, California 91125, USA}
	\affiliation{$^{4}$Institute of Microelectronics Technology and High Purity Materials RAS, Chernogolovka, Moscow Region, 142432, Russia}
	\affiliation{$^{5}$ Universit\'e Paris-Saclay, CNRS, Institut des Sciences Moléculaires d'Orsay, 91405 Orsay, France}
	\affiliation{$^{6}$Department of Condensed Matter Physics, Weizmann Institute of Science, Rehovot 76100, Israel}
    \affiliation{$^{7}$Dahlem Center for Complex Quantum Systems and Fachbereich Physik, Freie Universit\"at Berlin, 14195 Berlin, Germany}

\begin{abstract}
	
Second-order topological insulators are characterized by helical, non-spin-degenerate, one-dimensional states running along opposite crystal hinges, with no backscattering. Injecting superconducting pairs therefore entails splitting Cooper pairs into two families of helical Andreev states of opposite helicity, one at each hinge. Here we provide evidence for such separation via the measurement and analysis of switching supercurrent statistics of a crystalline nanoring of bismuth. Using a phenomenological model of two helical Andreev hinge modes, we find that pairs relax at a rate comparable to individual quasiparticles, in contrast with the much faster pair relaxation of non-topological systems. This constitutes a unique tell-tale sign of the spatial separation of topological helical hinges.

\end{abstract}
	\maketitle

Soon after the discovery of one-dimensional (1D) helical states in two-dimensional TIs (2DTI) \cite{Kane2005,Bernevig2006} or three-dimensional Second Order TIs (SOTI) \cite{Schindler2018a,Song2017,Langbehn2017}, it was realized that Josephson junctions containing helical modes as their weak link should display remarkable features. Indeed, the spin-momentum locking which characterizes the helical states translates into a fixed helicity for the Andreev states shuttling the supercurrent along each edge, in contrast to the spin degeneracy of conventional Josephson junctions. Among the predicted consequences are $4\pi$ \cite{Kwon04,FuKane2009} and $8\pi$ \cite{Zhang2014,Peng2016a} periodicities of the supercurrent-versus-phase relation (CPR) of a Josephson junction formed with a single helical edge state. Originating from fermion-parity protected crossings of  Andreev levels at phase difference~$\pi$,  these periodicities are contingent on the absence of fermion-parity-breaking processes. The necessity to beat such relaxation processes motivated the initial search for topological signatures at finite frequencies.

Past measurements have relied on the ac Josephson effect, via Shapiro steps \cite{Bocquillon2017} and Josephson emission of voltage-biased junctions \cite{Deacon2017}, or, as suggested in \cite{FuKane2009}, on the high-frequency response of a phase-biased topological junction \cite{Murani2019}.
Recent theoretical predictions suggest that signatures of topological superconductivity can also be found in switching current experiments conducted at frequencies comparable to the relaxation rate \cite{Yacoby2014,Peng2016,Recher2020,Crepin2014}. The idea is that the current at which the junction switches to its resistive state depends on the number and occupation of the current-carrying Andreev states. This implies that detailed information about the Andreev states and relaxation processes can be extracted from phase-dependent statistical distributions of switching currents \cite{Zgirski2011, Peng2016}. The particular sawtooth-like shape of the CPR makes long Josephson junctions with multiple subgap Andreev levels especially well suited for such investigations \cite{Beenakker2013,Recher2020}.

In this article we report measurements of the switching current distribution of a micrometer-size, ring-shaped bismuth monocrystal with superconducting contacts (see 1(a)). It was recently suggested
 \cite{Schindler2018} and supported by experiments \cite{Chuan2014,Yazdani2014,Takayama2015,Murani2017,Murani2019,Yazdani2019,Beidenkopf2019,Aggarwal2021} that Bi is a SOTI with helical modes propagating along its hinges. We find that in our device, the Bi ring acts as an intrinsically asymmetric Superconducting Quantum Interference Device (SQUID)  whose average switching current yields the characteristic sawtooth CPR of a long ballistic junction. In addition we show, by careful comparison to a phenomenological model, that the observed switching current behavior corroborates the existence of helical hinge modes in Bi. Our analysis leads to the identification of  single-particle and two-particle relaxation times,  both of the order of milliseconds,  consistent with well-separated topological hinge modes.

\begin{figure*}
	\centering
	\includegraphics[width=.9\textwidth]{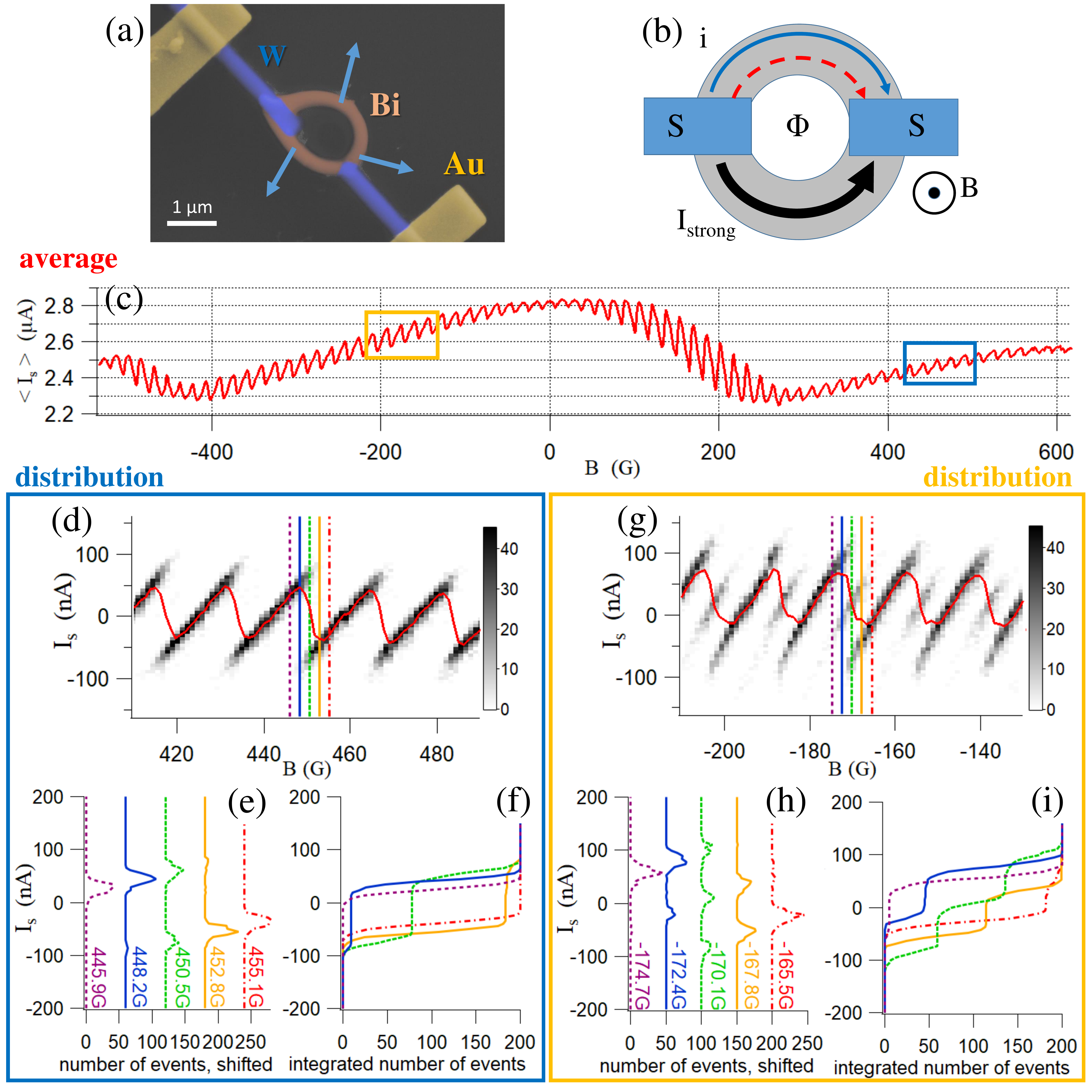}            
	\caption{\label{figaverage}
		Intrinsic asymmetric bismuth SQUID: Average switching current versus switching statistics. (a) Scanning electron micrograph with false colors of the Bi ring (brown) with superconducting W contacts (blue) and Au leads (yellow). The crystalline [111] axis (blue arrows) was found to keep a radial orientation (Methods). 	(b) Sketch of the bismuth nanoring connected to two superconducting contacts (S), forming an intrinsic Superconducting Quantum Interference Device (SQUID). In the model (see text), the total supercurrent splits into a small supercurrent $i$ carried by two helical channels (solid blue and dashed red arrows) in one branch of the ring, and a high supercurrent $I_\mathrm{strong}$ in the other branch. The magnetic field $B$ induces a flux $\Phi$ through the loop.	(c) Average switching current $<I_s>$  measured with a bias current ramp of $17~ \mathrm{Hz}$, as a function of $B$. $<I_s>$ displays a somewhat rounded sawtooth-shaped modulation with both signs of skewness, as well as regions (eg around zero field) where the modulation is more symmetric than a sawtooth. Note the asymmetry with field, related to the Josephson diode effect (see Supplementary Fig. 3 \cite{SM}). 	(d) and (g) Comparison of average (red lines) and full distribution (shades of grey) of switching current in two field ranges, $450~G$ (d) and $-170~G$ (g) for a 17 Hz current ramp. The baseline due to $I_\mathrm{strong}$ has been subtracted. The average switching current curves are rounded. By contrast, in (d) the distribution contains no switching events in the sawtooth discontinuities. Interestingly, there are regions around the discontinuities where two different switching currents are possible, as best seen in the histograms (e) and integrated histograms (f). Around  $-170~G$ (g), the switching current distributions display intermediate fainter branches, leading to some field regions where three values of the switching current are possible (see (h),(i)). In the text we argue that all features are well explained by the field-dependent occupation of Andreev states of two supercurrent-carrying helical hinges in the weak Josephson branch of the ring. 
	}
\end{figure*}

The average switching current at low fields, shown in Fig. 1 (c), displays periodic oscillations superimposed on a slowly varying baseline (see Methods and Supplementary Figures 1-4 \cite{SM}). The 17 G period, corresponding to one superconducting flux quantum $\Phi_0=h/2e$ through an area of $1.2~\mu m^2$, is consistent with the ring area. The oscillations have a (somewhat rounded) sawtooth shape, reminiscent of switching experiments on asymmetric SQUIDs
designed to measure the CPR of small Bi nanowire junctions \cite{Murani2017}. The sawtooth modulation corresponds to the CPR of a long ballistic Josephson junction, and thus demonstrated the higher order topological nature  of the Bi nanowire \cite{Cayssol2003,Beenakker2013,Murani2017,Schindler2018}.
In the present experiment, the sawtooth modulation suggests that the bismuth ring,  with its two superconducting contacts, acts {\it intrinsically} as an asymmetric SQUID, yielding a ballistic CPR for the branch of the ring with the smaller critical current. 

Rather than the average switching current, this article focuses on the switching current distribution, arguably a much more powerful (and underexploited) tool. We show that the distribution reveals the phase-dependence of the ground and excited states of the Andreev spectrum, their occupation probability and spatial separation, and hence their topological character. Two such distributions, recorded in two magnetic field regions, are displayed in Fig.~1(d) and (g). In contrast to the average, the switching current distributions are not rounded as a function of field. In the first magnetic field region, around  $B=450~G$, a notable feature of the sawtooth jump region are the two well separated peaks in the histogram (green curve in (1(e))). This indicates that the weak junction can be in two different states on the timescale of the current ramp.  In a second field region, near $B=-170~G$ (Fig. 1(g)), an additional intermediate, fainter branch develops around the sawtooth jump, so that there are three well separated peaks in the switching histograms (green curve in 1(h)). Correspondingly, the integrated switching current distributions display one (1(f)) or two (1(i)) intermediate plateaux.
In the following we argue that each peak in the histogram corresponds to a different occupation of the Andreev spectrum of a Josephson junction made of two helical hinges. Our analysis then yields the relative relaxation rates of the Andreev states, providing information about the topological character of these hinges. 

\begin{figure}
	\includegraphics[width=\columnwidth]{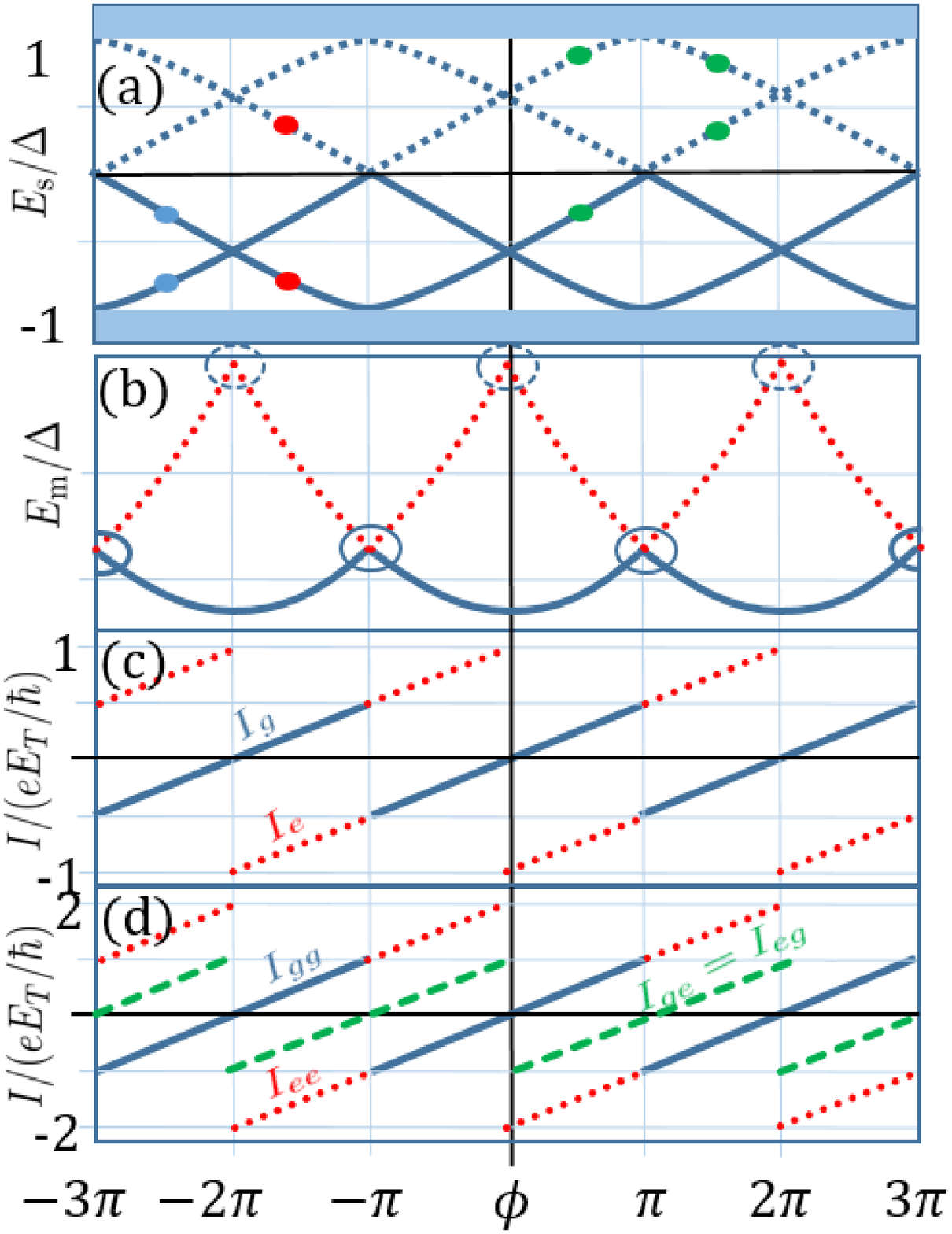}
	\caption{\label{Fig_CPRs} 
	Andreev spectrum and Josephson current of an intermediate-length Josephson junction between superconducting electrodes with pair potential $\Delta$ and phase difference $\phi$: case of one or two helical hinges in the junction. (a)~Andreev spectrum of single-particle (Bogoliubov-de Gennes) excitations $E_s$ for a single hinge mode. In the ground state, the two negative-energy states are occupied (blue circles). The lowest positive-energy state is occupied in the first excited state (and thus the corresponding negative-energy state empty, see red circles). Higher-energy excited states are indicated by green circles (but not included in our theoretical model). (b)~Andreev spectrum of corresponding many body states $E_m$ (given by the sum of Andreev levels and the continuum), including the ground state $E_g(\phi)$ (solid blue line) and the first excited state $E_e(\phi)$ (dashed red line). The excitation energy $\delta_E(\phi)=E_e(\phi)-E_g(\phi)$ is indicated by an arrow. Level crossings at odd multiples of $\pi$ (full-line circles) are protected by fermion parity, while level crossings at even multiples of $\pi$ (dashed circles) are protected by time reversal symmetry (strictly speaking broken by the magnetic field in the experiment).  (c)~Corresponding Josephson currents $i_g(\phi)$ in the ground state (solid blue line) and $i_e(\phi)$ in the first excited state (dashed red line), given by the derivative of the many-body energies with respect to $\phi$. $i_g$ is linear between $-\pi$ and $\pi$, with downward jumps by $ev_F/L\equiv \frac{e}{\hbar} E_T$, with $E_T=\hbar v_F/L$ the Thouless energy, at $\phi=\pi+2\pi n$ ($n\in\mathbb{Z}$), with $v_F$ the velocity of the hinge mode and $L$ the distance between the two superconductors. For the excited state, $i_e$ is also linear, with downward jumps by $2ev_F/L$ at $\phi=2\pi n$ and upward jumps by $ev_F/L$ at $\phi=\pi+2\pi n$. (d)~Josephson currents of a junction with two (identical) hinge modes. The current equals $i_{gg}(\phi)=2i_g(\phi)$ when both hinges are in their ground states, $i_{eg}(\phi)=i_{ge}(\phi)=i_g(\phi)+i_e(\phi)$ when one hinge is in the excited state, and $i_{ee}(\phi)=2i_e(\phi)$ when both are excited. }
\end{figure}

\begin{figure}
	\includegraphics[width=0.9\columnwidth]{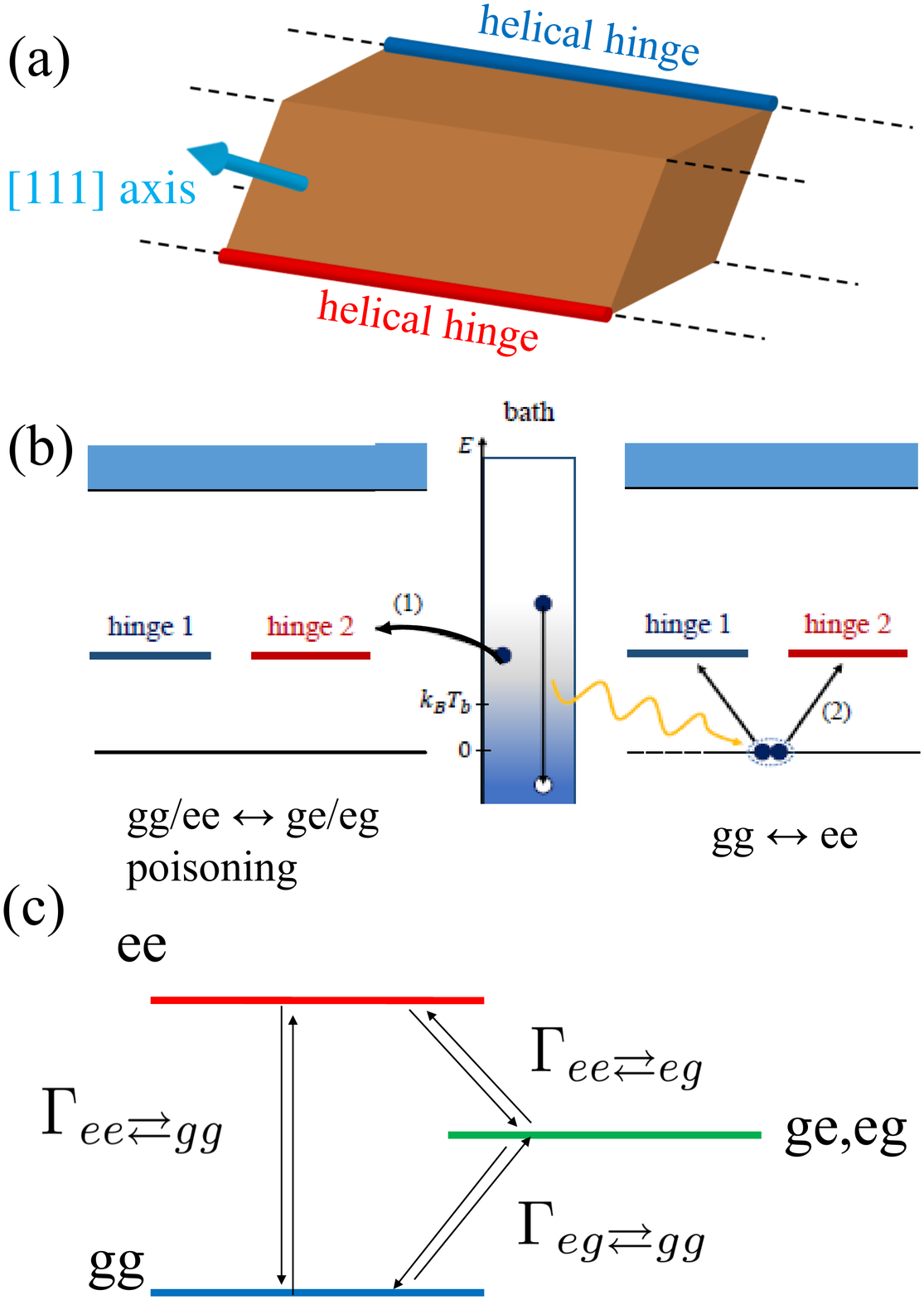}
	\caption{\label{Fig_transitions}  Excitation processes in the case of two spatially separated helical Andreev hinge states. (a) Sketch of a segment of the ring with a radial [111] axis (light blue arrow) and two helical hinge channels of opposite helicities on separate hinges (red and blue lines). (b) Quasiparticle and pair excitation processes. The dashed horizontal lines at zero energy represent the superconducting ground state, that acts as a Cooper pair reservoir. The blue region above energy $\Delta$ represent the quasiparticle continuum. The dark blue and red horizontal lines represent in-gap non-spin-degenerate helical Andreev bound states energy levels at a fixed $\phi$, each associated to a single helical (hinge) mode. Arrows represent processes that transfer one-particle occupation between two states. The intra-hinge or poisoning process (1) involves only one hinge and the quasiparticle bath, and the exchange of a quasiparticle that changes the parity of the hinge \cite{Devoret2018, Yeyati2014, Zgirski2011}. The inter-hinge or pair process (2) involves two hinges, an energy $2\delta_E$ from the bath, and a Cooper pair from the superconducting condensate. It does not change the global parity of the two-hinge system. This process is slower in the case of two separate helical hinges than in the case of one spin-degenerate non-helical channel. (c) Andreev energy levels of the two-hinge junction and corresponding relaxation rates. }
\end{figure}

The analysis is based on a model  of the Bi ring connected to two superconducting contacts, as a SOTI-based asymmetric SQUID: In the model, the weak Josephson junction consists of two helical supercurrent-carrying hinges located in one branch of the ring, and the strong junction, with higher critical current, is formed by the other branch (Fig. 1(b)). 
To leading order, the current ramp $\mathcal{I}(t)$ controls the phase difference $\gamma(t)$ across the strong junction: as $\mathcal{I}$ increases from zero to values close to the strong junction's critical current $I_{c,{\rm strong}}$, $\gamma$ increases from zero to $\gamma_\mathrm{max}$. Due to the flux threading the SQUID $\Phi$ (in units of $\hbar/(2e)$), the phase difference across the weak junction is $\phi(t)=\Phi +\gamma(t)$. The additional current through the weak junction with CPR $i(\phi)$ modulates the critical current at which the SQUID switches to a resistive state: $I_c \simeq I_{c,{\rm strong}} + i(\Phi+\gamma_\mathrm{max})$.  The SQUID's switching current thus provides a direct measurement of the current-phase relation (CPR) of the weak junction \cite{Zgirski2011, Peng2016, Delagrange2015}.

The CPR $i(\phi)$ reflects the  Andreev spectrum and, importantly, its occupations as a function of the  phase difference $\phi$~\cite{Beenakker2013}. Let us first consider the case of a junction made of a single hinge, as sketched in Figs 2(a-c): Panels 2(a) and 2(b) display the single-particle and many-body Andreev spectra, respectively. The many-body spectrum exhibits level crossings at integer multiples of $\pi$. The crossings at odd multiples are protected by fermion parity conservation, while those at even multiples require time-reversal symmetry.  Fermion-parity-violating processes are suppressed if the hinge mode is sufficiently isolated from other hinge modes or single-electron impurity states. To simplify, we focus on the two lowest-energy many-body states, whose CPRs, respectively $i_g(\phi)$ for the ground state and $i_e(\phi)$ for the excited state, are sketched in 2(c).  Both are piecewise linear and $2\pi$-periodic functions \cite{Beenakker2013,Crepin2014,Crepin2016}.
We now consider the case of two hinges. Given that each hinge can be either in the ground ($g$) or excited ($e$) state, there are four possible states for the system,  $gg$,$ee$ and $eg/ge$, whose CPR is $i_{ll'}=i_l(\phi) + i_{l'}(\phi)$, with $l,l'$ (g) or (e), see 2(d). In the case of two hinge channels in the long junction regime, with same critical currents,  $i_{gg}=2i_g$, $i_{ee}=2i_e$, and $i_{ge}(\phi)=i_{eg}(\phi)=i_g(\phi) + i_{e}(\phi)=i_{gg}(\phi+\pi)$, which is the sawtooth-shaped CPR of the ground state shifted by $\pi$. The three different CPRs are sketched in 2(d).

In the following we use this model to compute the occupation probability of the various Andreev levels that matches the one extracted from the experimental switching current distribution. 
To this end, we first solve the rate equations for the probabilities $p_{gg}$, $p_{ge}=p_{eg}$, and $p_{ee}$ of occupying states $gg$,$ee$ and $eg/ge$. The rate equations are:
 \begin{gather}
 \frac{dp_{gg}}{dt}	=-2\Gamma_{eg\leftarrow gg}p_{gg}+2\Gamma_{gg\leftarrow eg}p_{eg}-\Gamma_{ee\leftarrow gg}p_{gg} \nonumber\\
 +\Gamma_{gg\leftarrow ee}p_{ee} \nonumber\\
 \frac{dp_{eg}}{dt}	=-\Gamma_{gg\leftarrow eg}p_{eg}+\Gamma_{eg\leftarrow gg}p_{gg}-\Gamma_{ee\leftarrow eg}p_{eg} \nonumber \\
 +\Gamma_{eg\leftarrow ee}p_{ee},
 \label{eq:ratetwohinge}
 \end{gather}
 with $p_{ee}=1-2p_{eg}-p_{gg}$.
 They include two types of relaxation processes, sketched in 3(b) and (c): The intra-hinge, or poisoning processes, cause one hinge to be excited or relax with a rate $\Gamma_{ee\rightleftarrows ge}$, involving a relaxation time $\tau_2$, or  $\Gamma_{eg\rightleftarrows gg}$, involving a relaxation time $\tau_1$. The inter-hinge  or pair processes with rates $\Gamma_{gg\rightleftarrows ee}$, in which two quasiparticles from different hinges condense into one Cooper pair (or a Cooper pair splits to populate the two hinges), involves a pair relaxation time $\tau_p$. Following \cite{Yacoby2014}, we assume that the intra-hinge transition rates involve a fermionic bath at a temperature $T_{\rm qp}$: $ \Gamma_{ee\rightleftarrows eg}(\phi) = f(\pm\delta_E(\phi)/k_B T_{\rm qp})/\tau_2$, with $f$ the Fermi distribution function, and $\delta_E(\phi) = E_e(\phi) - E_g(\phi)$ is the gap between the ground and excited states. 
Similarly,  $\Gamma_{eg\rightleftarrows gg}(\phi) = f(\pm\delta_E(\phi)/k_B T_{\rm qp})/\tau_1$. The interhinge or pair rates also involve a fermionic bath, at a temperature $T_{\rm b}$ (that can be different from $T_{\rm qp}$). These inter-hinge relaxation processes do not require external particles from the fermionic bath, but only energy. The rates thus contain the Bose-Einstein function and twice the excitation energy $2 \delta_E(\phi)$. They are suppressed for hinge modes that are far apart in real space on the scale of the superconducting coherence length.

\begin{equation}
\Gamma_{ee\rightleftarrows gg} = \frac{2\delta_E(\phi)}{E_T\tau_p}
\begin{cases}
 & n_B\left(\frac{2\delta_E(\phi)}{k_B T_b}\right) \text{ } \\
 & 1+ n_B\left(\frac{2\delta_E(\phi)}{k_B T_b}\right) 
\end{cases}
\end{equation}
where $E_T=\hbar v/L$ is the Thouless energy and $n_B(x) = \left(e^x - 1\right)^{-1}$ is the Bose function. The probabilities are obtained from numerically integrating Eq. (\ref{eq:ratetwohinge}) from $\phi =\Phi$ to $\phi_\mathrm{sw}=\Phi+\gamma_\mathrm{max}$, assuming that $\gamma(t)$ increases linearly in time, $\gamma(t)=\omega t$, from zero to $\gamma_\mathrm{max}$ as $\mathcal{I}(t)$ ramps up from zero to $I_c$. We choose $\gamma_\mathrm{max}=\pi/2$ which best fits the experimental data (see Supplementary Figure 5 for $\gamma_\mathrm{max}=\pi$ \cite{SM}). The initial conditions are the equilibrium probabilities, see details in Supplementary \cite{SM}. From the probabilities of occupying the different states we compute the probability to switch to a dissipative state. The switching current statistical distribution is then generated taking  into account the fact that for a given state switching is a stochastic event characterized by a current probability distribution. We thus introduce a state-dependent switching probability $P^{l,l^\prime}_\mathrm{sw}(I,\phi_\mathrm{sw})$, which is the probability of finding the SQUID in the resistive state at bias current $I$ and switching superconducting phase difference $\phi_\mathrm{sw}$, for a given occupied state $ll^\prime$. We approximate $P^{l,l^\prime}_\mathrm{sw}(I,\phi_\mathrm{sw})$ by a smoothed step function of width $\delta I$ centred around the SQUID's critical current $I_c^{ll^\prime} = I_c^l + I_c^{l^\prime}$ (which depends on the state of the weak junction through $i_{l,l^\prime}(\phi_\mathrm{sw})$). The total switching probability is then expressed as
$P(I,\phi) = \sum_{l,l^\prime\in\{e,g\}} p_{ll^\prime}(\phi) P_{\rm sw}^{ll^\prime}(I,\phi).$

To compare experiment and theory, we extract from the experimental switching current distributions the field-dependent histograms and integrated histograms, from which we derive the state-dependent experimental occupation probabilities (see methods). The theoretical occupation probabilities are then computed using the parameters $\omega\tau_1$,  $\omega\tau_2$, $\omega\tau_p$, $T_b$ and $T_{qp}$ which best reproduce the experimental occupation probabilities. $P_{\rm sw}^{l}(I,\phi_\mathrm{sw})$, $dP(I,\phi_\mathrm{sw})/dI$ and the full switching distribution as a function of flux are subsequently generated with those parameters. 

Fig. 4 displays how well the experimental switching current distribution around -170 G are reproduced by theory. Two current ramp frequencies, 17 and 187 Hz,  were investigated. The model reproduces the extent over which the fainter intermediate (poisoning) branch extends, and how it extends further in the case of the higher current ramp frequency. 
The model also reproduces remarkably well the shape, height, and relative positions of the three probability distributions $p_{gg}$, $p_{eg}+p_{ge}$ and $p_{ee}$ extracted from the integrated experimental histogram (compare  3(b) with (f) and (d) with (h)). In the regions with three possible switching currents, there are three non negligible occupation probabilities of the states $gg$, $ee$ and $eg$. For the slowest ramp, $p_{gg}$ and $p_{ee}$ are extremal at  $\pi$, whereas $p_{eg}+p_{ge}$ is maximal slightly above $\pi$ (3(d)). The corresponding plot at a ramp frequency eleven times greater, 3(h), displays a much greater shift of the maximum of $p_{eg}+p_{ge}$. This shift is the signature of the inter-hinge pair relaxation processes of typical time $\tau_p$. The parameters used are $\tau_{qp}=\tau_1=\tau_2=10.5~\rm{ms}$ and $\tau_p=1.82~\rm{ms}$ for both the slow and fast ramp. Only $T_b=T_{qp}$ was allowed to change, yielding $k_BT_{\rm b}/E_T\simeq 0.4$ for 17 Hz  and $k_BT_{\rm b}/E_T\simeq 0.7$ for 187 Hz,  reflecting the smaller time available for quasiparticle thermalisation in the reservoirs. 
We estimate a factor two uncertainty for $\tau_{qp}$ and five for $\tau_p$, and thus a factor seven uncertainty for the ratio $\tau_{qp}/ \tau_p$ \cite{SM} (See Supplementary figures 5 and 6 for the effects of the different parameters and the precision with which they can be estimated).

We now turn to modelling the experimental switching current distribution around 450 G, see Fig. 5. Interestingly, in this field region hardly any intermediate switching branch is visible (5(a) and (c)). This means that the $ee$ and $gg$ states are much more populated than the $eg$ state, as clearly seen in the extracted occupation probabilities, 5(b) and (d): the poisoned state probability $p_{eg}$ is less than $5\%$, with a maximum shifted with respect to the $p_{ee}$ and $p_{gg}$ extrema. This situation, with very little poisoning, is unexpected since it corresponds to a higher probability of the more energetic $ee$ state than the $eg$ state. It can be reproduced using a slow pair relaxation time and a relaxation time $\tau_2$ out of the $ee$ state and into the $eg$ state that is ten time longer than the relaxation time $\tau_1$ out of $eg$ and into $gg$ (see 3(c)). In addition, a much smaller quasiparticle temperature is required compared to the pair bath temperature, along with a small gap in the Andreev spectrum (too small to be detected in the experiment). The parameters are the same $\tau_p=1.82~ms$ pair relaxation time as previously, but $\tau_1=25~ms$ and $\tau_2=250~ms$. Given the larger number of parameters involved, we consider the description of this low poisoning regime qualitative rather than quantitative. (See Supplementary Figures 7 and 8 for the comparison of experimental and theoretical switching current histograms, integrated histograms, and probability distributions, for both field regions and both sweep rates).

 \begin{figure*}
	\includegraphics[width=0.9\textwidth]{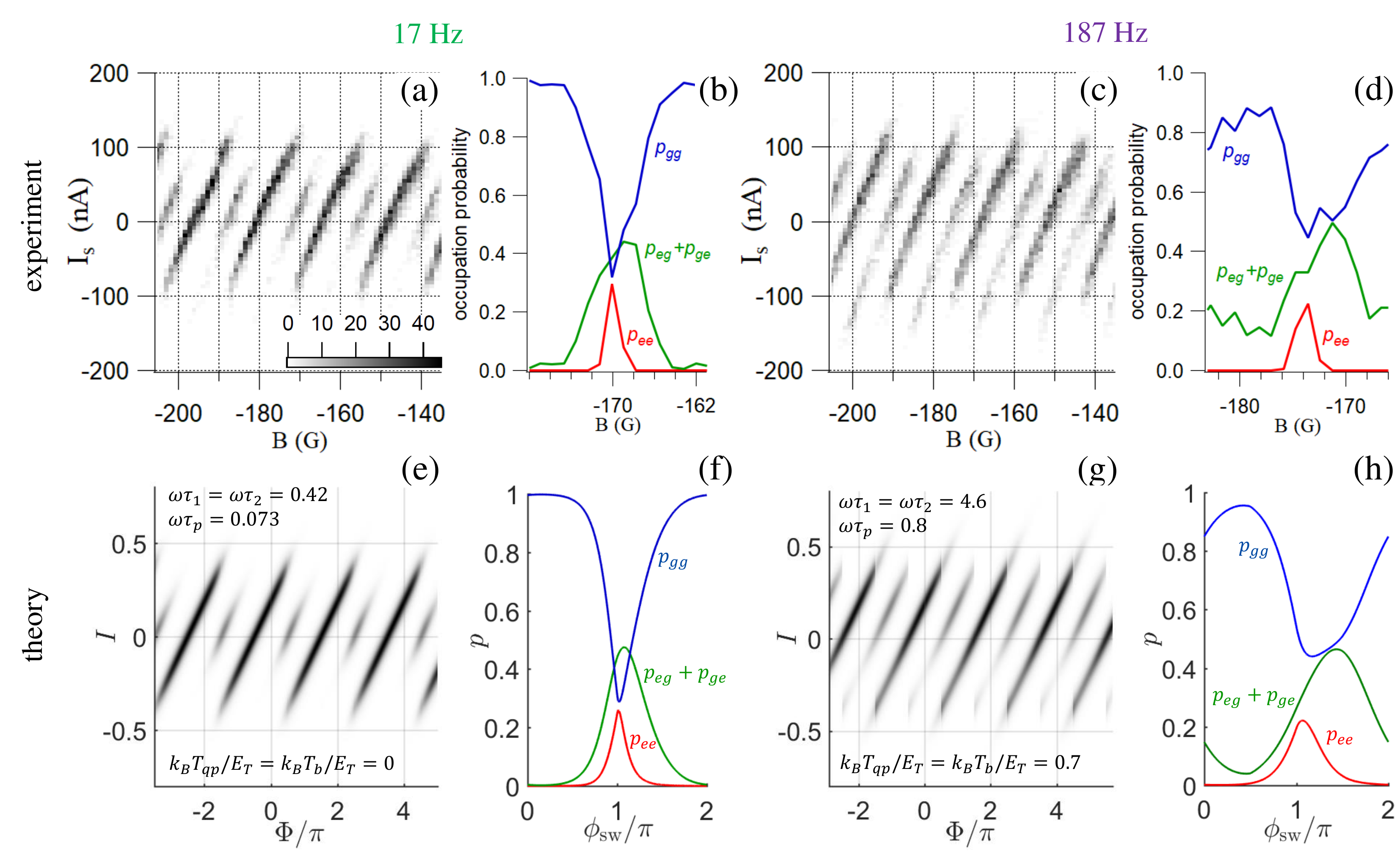}
	\caption{\label{Figm170G} Measured switching current distributions and extracted (see Methods) Andreev hinge states occupations probabilities around $-170~G$, for a ramp frequency of 17 Hz and 187 Hz (top); Comparison to theory (bottom). 	(a) (respectively (c)): Switching current histograms over four flux periods at 17 Hz (respectively 187 Hz)). The number of switching events is coded in shades of grey. A linear dependence has been subtracted to remove the critical current of the strong junction. (b) (respectively (d)): Field-dependence  of the occupation probability of three configurations, corresponding respectively to both hinges in the ground state ($p_{gg}$, blue), both hinges in the excited state ($p_{ee}$, red), or one hinge in the ground state and the other in the excited state ($p_{eg}+p_{ge}$, green), at 17 Hz (respectively 187 Hz). The corresponding theoretical curves are computed (see Methods) using for both frequencies the same relaxations times $\tau_{qp}$ and $\tau_p$ (and same Thouless energy). Only the bath temperature was allowed to vary. The Thouless energy $E_T=\hbar v_F/L\simeq 1.5~k_B K$, taking Fermi velocity $v_F\simeq 4.10^5~\rm{m/s}$ and junction length $L \simeq 2~\rm{\mu m}$.  (e) and (f) are computed with $k_BT_b/E_T=k_BT_{qp}/E_T=0.4,~ \omega \tau_1=\omega \tau_2=0.42, ~\omega \tau_p=0.073$. (g) and (h) are computed with  $k_BT_b/E_T=k_BT_{qp}/E_T=0.7, ~\omega \tau_1=\omega \tau_2=4.6,~ \omega \tau_p=0.8$. The corresponding times are $\tau_{qp}=10.7~\rm{ms}$ and $\tau_p=1.9~\rm{ms}$.
	The theory reproduces quite well the intermediate switching branch and its extension with increasing ramp frequency, corresponding to a shift in phase of $p_{eg}$. All occupation probabilities are also well reproduced. The asymmetry of the switching current distribution, reflecting the finite relaxation times, is visible in (a) and (c) thanks to the intermediate distribution. Correspondingly, the asymmetric shapes of the occupation probability peaks (or dips) are clearly visible in experiment ((b),(d)) and theory ((f),(h)). The model fails however to capture some experimental features at $187~\rm{Hz}$: In experiment (c), the main branch is asymmetric towards positive current, whereas the intermediate, fainter branch is asymmetric  towards negative current. In the switching statistics generated by theory, by contrast (g), the main branch extends further, for both positive and negative current, than the intermediate branch. This discrepancy may be attributed to our restricting the model to the first excited state only, see Fig.~2(b).}
\end{figure*}

\begin{figure*}[t]
\includegraphics[width=\textwidth]{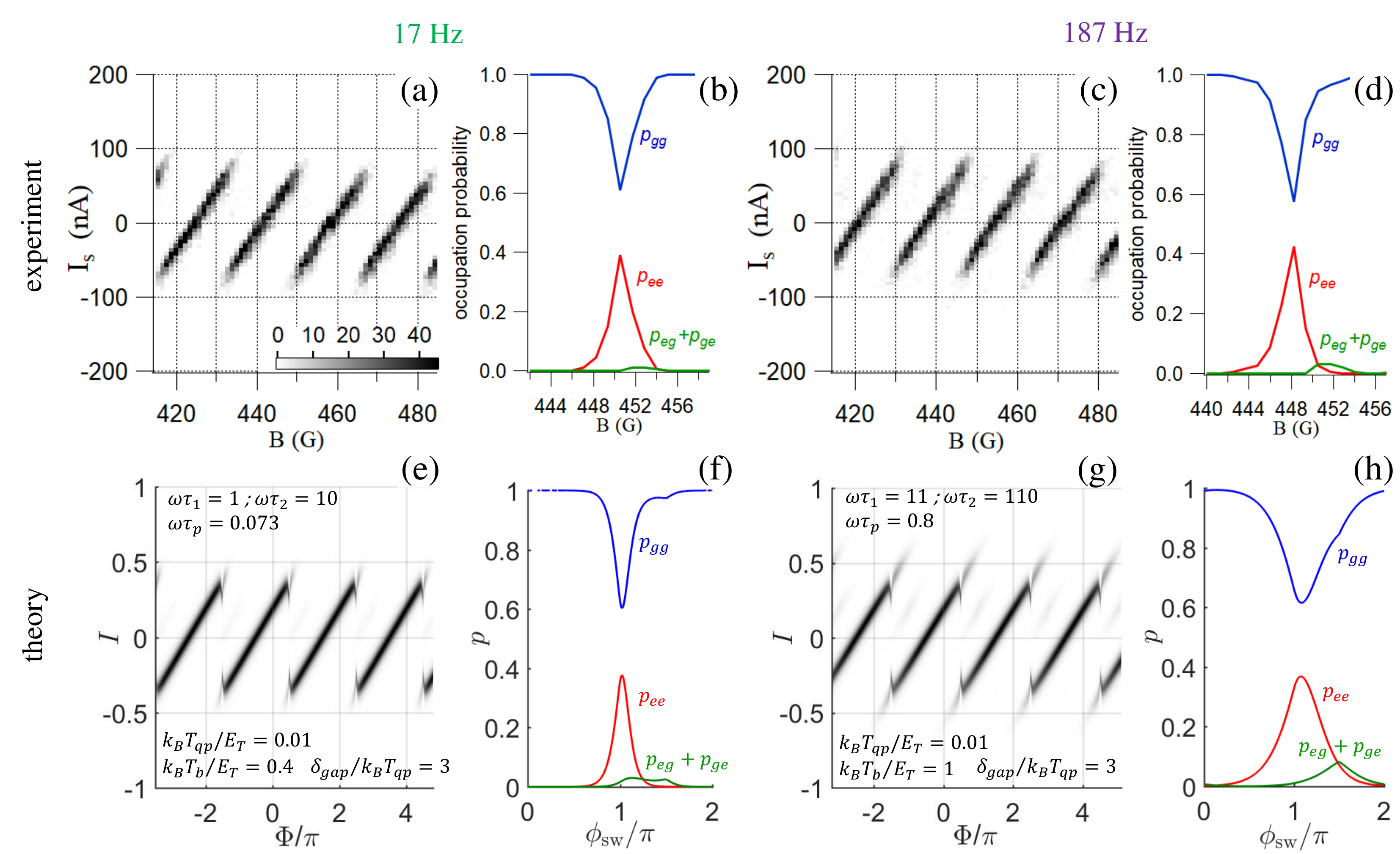}
\caption{\label{Fig450G}
 Comparison of the experimental switching statistics and extracted Andreev hinge states occupation probabilities to the results of the theoretical model in a field region centered at $450~G$, with very little poisoning. (a) and (c) Switching current histograms over four flux periods around 450 G,  for a current ramp frequency of 17 Hz adn 187 Hz, respectively. The number of switching events is coded in shades of grey. A linear dependence has been subtracted to remove the critical current of the strong junction. The intermediate switching current branch is practically not visible. (b) and (d) Field-dependence of the occupation probability of the Andreev hinge states, extracted from the integrated distributions (see e.g. Fig 1(e)). The occupation probabilities are almost entirely distributed between the ground ($p_{gg}$, blue curve) and excited ($p_{ee}$, red curve) states. The probability of occupying the eg and ge states is less than $5\% $. The corresponding theoretical curves, shown in (e), (f), (g) and (h), computed including a small gap in the spectrum, and two different relaxation times for quasiparticles (see text) reproduce the qualitative features of this small-poisoning regime, including slight increased visibility of the poisoned state at higher sweep rate. The parameters are $k_BT_b/E_T=0.4$, $k_BT_{qp}/E_T=0.01$, $\omega \tau_p=0.073$, $\omega \tau_1=1$ and $\omega \tau_2=10$ at 17 Hz, and $k_BT_b/E_T=1$, $k_BT_{qp}/E_T=0.01$, $\omega \tau_p=0.8$, $\omega \tau_1=11$ and $\omega \tau_2=110$ at 187 Hz, yielding for the  
 relaxation times $ \tau_p=1.82~\rm{ms}$, $\tau_1=25~\rm{ms}$ and $\tau_2=250~\rm{ms}$. The gap in the spectrum at $\pi$ is $3k_BT_{qp}$.} 
\end{figure*}

Our analysis has led to the identification of three times, describing respectively the intra-hinge relaxation from the excited  to the ground state within a single hinge  (single-quasiparticle  or poisoning process, with times $\tau_1$ and $\tau_2$), and the inter-hinge or pair relaxation  involving a two-particle process in which two hinges simultaneously acquire or release a quasiparticle over a time~$\tau_p$. This process is impeded if the hinges are far apart, and correspondingly the time  $\tau_p$ should increase with the separation between hinges. 
Let us compare the values of $\tau_p$ and $\tau_{qp}$ we have found (albeit overestimated because of possible inductance effects, see Supplementary Figures 9 and 10 \cite{SM}), in the $\sim 10-100~\rm{ms}$ range, to the values obtained in nontopological junctions in similar environments. The poisoning relaxation times we find are similar to the ones measured in Josephson junctions based on atomic contacts \cite{Zgirski2011,Yeyati2014} and semiconducting nanowires \cite{Devoret2018,Devoret2021}, which vary between a few hundred $\mu s$ and $ms$. In striking contrast, the pair relaxation times $\tau_p$ estimated in those works, and associated to the $T_1$ relaxation time of the Andreev qubit \cite{Janvier2015} are two to three orders of magnitude shorter, in the $\mu s$ range, than what we find in the bismuth nanowire.
We interpret this as demonstrating the strong decoupling between hinges, confirming the topological character of bismuth. Indeed, while in  a nontopological Josephson junction, every helical channel locally coexists with its opposite helicity counterpart, in a topological system, the two helical channels are spatially separated, typically by one hundred nanometers or more. This separation is roughly one hundred times greater than the transverse extension of the helical Andreev states at the Bi nanowire hinges 
 \cite{Murani2017,Chuan2014,SM}), and ten times greater than the superconducting coherence length of the disordered W contacts (typically a few nanometers).
A remaining puzzle is why the degree of poisoning depends on magnetic field, so that poisoning is clear in one field range and practically undetectable in another.  One possibility is that the Zeeman field, by tilting the spins, can remove the orthogonality between states of a given hinge, thereby allowing spin-conserving, backscattering relaxation/poisoning transitions within one hinge. Whereas when the states are orthogonal, backscattering relaxation must occur through a change of hinge, which is very slow if the hinges are separated. A second possibility, mentioned in \cite{Yacoby2014}, explains the change in effective temperature of the poisoning quasiparticles $T_{qp}$ by a change in the nature and number of quasiparticles that can couple to Andreev bound states. Depending on magnetic field, the Andreev bound states, whose energy shifts with Zeeman field, could be alternately coupled to the quasiparticle continuum above the superconducting gap (yielding a rather large temperature), or coupled only to rarer localized states (corresponding to a very small temperature). We believe that both the number of available quasiparticles and the selection rules given by the helical nature of the hinges could explain the two very different regimes we see.

In conclusion, our investigation of the full switching current statistics of a bismuth nanoring Josephson junction provides an unprecedented look into Second Order Topological Insulators and the helical Andreev Bound States that are predicted to carry the supercurrent along spatially separated 1D hinges. 
Our detection around phase $\pi$ of switching events originating from both excited and ground states, on millisecond timescales, is a demonstration of slow relaxation of quasiparticles and, more spectacularly, of pairs. These features are an unambiguous signature of the topological protection provided by parity conservation in Quantum Spin Hall state-based Josephson junctions, opening new possibilities for the design of protected qubits \cite{Klinovaja2018}. In addition, the SOTI hypothesis of transport occurring through two Andreev hinge states of opposite helicities situated at two separate hinges is confirmed by the unusually long pair relaxation time compared that of non-topological materials.
We believe the full statistical measurement of the switching current is a simple yet powerful technique that will prove useful to investigate topology and correlations between current-carrying paths in a vast range of Josephson junctions, in particular based on 2DTI and other SOTI materials, such as WTe2 \cite{Kononov} and $\mathrm{Cd_3As_2}$ \cite{ChuanPRL2020}.

\section{Acknowledgments}

We acknowledge useful discussions with Marco Aprili, Caglar Girit, Manuel Houzet, Julia Meyer,  Hugues Pothier, Pascal Simon, Cristian Urbina,  technical help from S. Autier-Laurent, and funding from the French program ANR JETS (ANR-16-CE30-0029-01), funding from the European Research Council (ERC) under the European Union's Horizon 2020 research and innovation programme (grant Ballistop agreement n°  833350), LabEx PALM (ANR-10-LABX-0039-PALM) JosephBismuth, CRC 183 (project C02) of Deutsche Forschungsgemeinschaft (YO and FvO), European Union's Horizon 2020 research and innovation programme (Grant Agreement LEGOTOP No. 788715) (YO), ISF Quantum Science and Technology (2074/19) (YO), and the BSF and NSF grant (2018643) (YO). YP is supported by the NSF grant (PHY-2216774) and the startup fund from California State University, Northridge.

\section{Author Contributions}
 AK and VV grew the bismuth nanowires and AK deposited them on the substrate, YuK characterized the nanowire growth.  FF, AB and AK selected nanowires and connected them with Focused Ion Beam-assisted deposition. AB conducted the low temperature measurements with input from MF, RD, SG and HB. AB, MF, RD, SG, HB, YP, FvO and YO analysed the data and discussed the results. YP, FvO and YO developed the model and YP wrote the  code. SG, HB, AB, RD, YP, FvO and YO wrote the paper.

\section{Competing Interest}
The authors declare no competing interests.

\clearpage

\section{Methods}
\subsection{Bismuth nanoring-based SQUID}
Low defect, monocrystalline bismuth nanowires were grown by sputtering high purity bismuth onto a Si substrate covered by a thin layer of vanadium ($T\simeq70^\circ {\mathrm C}$). The shock wave of short laser pulses was used to shake off nanowires, as in Ref.~\cite{Kasumov1999}, and transfer them contactlessly onto a substrate with prepatterned markers. A few nanowires coil into rings during the transfer. We selected the loop-shaped bismuth nanowire shown in Fig.~1(a), and followed its crystalline orientation at several points along the ring using Electron Backscatter Diffraction (EBSD). As represented by the light blue arrows in Fig.~1(a), the [111] crystal axis rotates along the ring, in an almost radial orientation. An idealized section of the ring is sketched in Fig.~3(a), with the helical hinge channels characteristic of SOTIs. The ring was contacted using gallium Focused-Ion-Beam-assisted deposition of a superconducting tungsten compound, after a step of Ga etching to remove the oxide layer covering the bismuth surface. The tungsten compound is a disordered superconductor, with a gap $\Delta \sim 1~\mathrm{meV}$ and a critical field higher than 8 T. Based on a careful analysis of several samples using Energy Dispersive Spectrocopy and etching, we can assert that tungsten contamination extends less than $d \simeq 300 ~ \mathrm{nm}$ around the deposition regions. The tungsten contacts were connected to thick titanium-gold electrodes.
\subsection{Switching current measurements} Measurements were carried out in a dilution refrigerator with a base temperature of $100 ~\mathrm{mK}$ via low-pass filtered lines and RC filters of cut-off frequency~$\sim10~ \mathrm{kHz}$. A magnetic field of up to $12~\mathrm{T}$ could be applied perpendicular to the sample plane. The switching current was measured using a counter synchronized with a current ramp of frequency $17 \rm{~or}~187 ~\mathrm{Hz}$, triggered by a voltage jump each time the system switches from the supercurrent-carrying to the resistive state (see \cite{SM} for more details).  $250$ (respectively $200$) switching events were recorded for each value of magnetic field,  to measure the average (respectively full distribution of the) switching current. 
\subsection{From experimental supercurrent switching statistics to occupation probabilities}
The occupation probabilities are extracted from the integrated switching current distributions (see e.g. Fig. 1(f) or 1(i)) by noting that each step in the integrated distribution corresponds to a transition out of a specific supercurrent-carrying state. The height of the step from one plateau to the next therefore counts the number of switching events from that state, and is normalized to yield the occupation probability of that state just before the switching event. 
\subsection{Sequence of resolution of the phenomenological model }
The theoretical model first generates the occupation probabilities $p_{gg}$, $p_{eg}$ and $p_{ee}$ as a function of $\phi_{sw}=\Phi+\gamma_\mathrm{max}$, with $\gamma_\mathrm{max}=\pi/2$, see text. Those are displayed in panels (f) and (h) of  Figures 4 and 5. The switching current probability $P(I)$ and its derivative  $dP/dI$ are then generated (as shown in Supplementary figures 7 and 8), from which a switching current histogram is created (panels (e) and (g) of  Figures 4 and 5).  
\subsection{Data availability}
The switching current data are available upon request at https://doi.org/10.5281/zenodo.7119795
\subsection{Code availability}
 The matlab files for calculating the joint probabilities and switching histograms can be found at $\mathrm{https://github.com/pengyangraul/BiJunction_Codes}$

\clearpage

\end{document}